\documentstyle[12pt]{article}

\input epsf
\begin{document}
\begin{titlepage}
\today          \hfill
\begin{center}
\hfill    OITS-669 \\

\vskip .05in

{\large \bf Limits on Low Scale Gravity from $e^+ e^- \rightarrow W^+ W^-$, 
$ZZ$ and $\gamma \gamma$
}
\footnote{This work is supported by DOE Grant DE-FG03-96ER40969.}
\vskip .15in
K. Agashe \footnote{email: agashe@oregon.uoregon.edu} and
N.G. Deshpande \footnote{email: desh@oregon.uoregon.edu}
\vskip .1in
{\em
Institute of Theoretical Science \\
5203 University
of Oregon \\
Eugene OR 97403-5203}
\end{center}

\vskip .05in

\begin{abstract}
It has been proposed recently that the scale of quantum gravity 
(``the string scale'') 
can be
$M_S \sim$ few TeV with $n \geq 2$ extra dimensions of size $R 
\stackrel{<}{\sim}$ mm so 
that, at distances greater than $R$, Newtonian gravity with $M_{Pl} \sim
10^{18}$ GeV is reproduced if $M_{Pl}^2 \sim R^n M_S^{n+2}$.
Exchange of virtual gravitons in this theory generates higher-dimensional
operators involving SM fields, suppressed by powers of $M_S$.
We discuss constraints on this scenario from the contribution
of these operators to the processes $e^+ e^- 
\rightarrow W^+ W^-$, 
$ZZ$, $\gamma \gamma$. 
We find that LEP2 can place a limit $M_S \approx 1$ 
TeV from $e^+ e^- 
\rightarrow W^+ W^-$,
$ZZ$, $\gamma \gamma$.

\end{abstract}

\end{titlepage}

\newpage
\renewcommand{\thepage}{\arabic{page}}
\setcounter{page}{1}

\section{Introduction}
A new framework to solve the gauge hierarchy problem of the Standard 
Model (SM) has recently been proposed by Arkani-Hamed, Dimopoulos, Dvali 
\cite{add}: the scale at which the gravitational interactions become 
comparable in strength to the ordinary gauge interactions (``the 
string scale''), $M_S$, is close to the weak 
scale, $m_W$, {\it i.e.,} $\sim$
TeV. The ultraviolet cut-off for the quadratically 
divergent quantum corrections
to the Higgs (mass)$^2$ is then $\sim$ TeV, thus stabilizing the weak scale.
In other words, there
is no hierarchy to begin with since the only fundamental scale
is the weak scale $m_W \sim$ TeV. 

To get the Newtonian $1/r$ gravitational potential
with scale $M_{Pl} \sim 10^{18}$ GeV, it is proposed that there are 
$n$ extra dimensions of size $R$. The ``Planck scale'' in $(4 + n)$
dimensions is $\sim M_S \sim 1$ TeV, but the SM particles propagate only
in the usual $4$ dimensions while gravitons (and perhaps other particles)
propagate in $(4 + n)$
dimensions. Thus, for $r \ll R$,
the gravitational potential is modified form the usual one whereas
for $r \gg R$, using Gauss' law, we can see
that the Newtonian $1/r$ potential is recovered with the scale $M_{Pl}$
if the following relation between $n$ and $R$ is satisfied \cite{add}:
\begin{equation}
M_{Pl}^2 \sim R^n M_S^{n+2}.
\end{equation}
If $n = 1$, $R$ is too large and is ruled out since
gravity is then modified over solar system distance scales. For 
$n \geq 2$, we get $R \stackrel{<}{\sim}$ mm. Gravity 
has been measured at present only to
distance scales $\sim$ mm.

From the $(4 + n)$ dimensional point of view, the gravitons couple
to the SM fields with strength given by powers of
$(1/M_S)$, thus inducing (from graviton exchange)
higher-dimensional
operators involving SM fields, 
suppressed by powers of $M_S$.
Seen from the $4$ dimensional point of view, a graviton with momentum in the
$n$ compact dimensions (of size $R$) behaves as a particle with mass 
$\sim 1/R$.
Each graviton couples to the SM fields with strength suppressed by powers of 
the $4$ dimensional Planck scale, $M_{Pl}$. However, there is a large
multiplicity of these graviton states 
since we have to sum over all possible
momenta in the $n$ dimensions.
This tower of Kaluza-Klein graviton states results in an enhancement 
of this
coupling to powers of $(1/M_S)$ \cite{add}.

Since the gravitons couple to all SM particles, 
the $s$, $t$ and $u$-channel exchange of virtual gravitons mediates 
processes like $f \bar{f}$, $VV$ 
$\rightarrow f^{\prime} \bar{f}^{\prime}$, $V^{\prime} V^{\prime}$
(where $f$ are fermions and $V$ are
gauge bosons),
$e \; (\hbox{or}\; \nu) \; q \rightarrow e \; (\hbox{or}\;
\nu) \; q$ etc. \cite{grw,hlz,h,others}.

In this letter, we study the contribution of the virtual graviton
exchanges in these theories with quantum gravity at the weak scale to
the processes $e^+ e^- \rightarrow W^+ W^-$, $ZZ$,
$\gamma \gamma$ at LEP2 and at a future
Linear Collider (LC).

\section{Matrix Elements}
The matrix element for 
\begin{equation}
e^+ \; (p_+) \; e^- \; (p_-) \rightarrow V \; (k_1) \bar{V}_2 (k_2)
\end{equation}
from graviton exchange (neglecting the electron mass)
is given by \cite{grw,hlz} (we use Eqn.(71) of Han, Lykken, Zhang in
\cite{hlz}) 
\begin{eqnarray}
{\cal M}_{gravity} & = & \lambda
\frac{4}{M_S^4} \left[ 2 \left(p_- \cdot 
k_2 - p_- \cdot k_1 
\right) \left( \varepsilon^{\ast}_1 \cdot \varepsilon^{\ast}_2 \right) 
\bar{v} \not \! k_1 u 
\right.
\nonumber \\
 & & + 2 \left( p_- \cdot \varepsilon^{\ast}_1 \right) \left( k_1 \cdot 
\varepsilon^{\ast}_2 
\right) \bar{v} \not \! k_1 u - 2 \left( p_- \cdot \varepsilon^{\ast}_2 
\right) 
\left( k_2 \cdot \varepsilon^{\ast}_1 
\right) \bar{v} \not \! k_1 u  \nonumber \\
 & & -2 \left( p_- \cdot k_2 \right) \left( k_1 \cdot \varepsilon^{\ast}_2 
\right) \bar{v} \not \! \varepsilon^{\ast}_1 u
-2 \left( p_- \cdot k_1 \right) \left( k_2 \cdot \varepsilon^{\ast}_1 
\right) \bar{v} \not \! \varepsilon^{\ast}_2 u \nonumber \\
 & & \left.
+ s \left( p_- \cdot \varepsilon^{\ast}_2 \right) \bar{v} \not \! 
\varepsilon^{\ast}_1 u
+ s \left( p_- \cdot \varepsilon^{\ast}_1 \right) \bar{v} \not \! 
\varepsilon^{\ast}_2 u \right],  
\label{mgravity1}
\end{eqnarray}
where the momenta are of the incoming $e$'s and the outgoing bosons,
the $\varepsilon$'s are the polarization vectors of the gauge bosons and
$\bar{v}$, $u$ are the $e^+$, $e^-$ spinors,
$s$ and $t$ denote the Mandelstam variables. 
The scale $M_S$ is chosen to agree
wth the notation used by Hewett in \cite{h} (see
Eqns.(61) and (5) of \cite{hlz} and \cite{h}, respectively). The factor 
$\lambda$ as in \cite{h} incorporates
any model-dependence ({\it i.e.,} it depends on the full theory -- we
will assume that it is $\pm 1$). So, strictly speaking, our limits
are on $\mid \lambda \mid ^{-1/4} M_S$. \footnote{It is possible that
(depending on the full theory)
other tree level diagrams or loop diagrams generate other 
higher-dimensional
operators which may also contribute to the process
$e^+ e^- \rightarrow V \bar{V}$. In this case, the limits on
$M_S$ will be modified.} 

The helicity amplitudes, ${\cal M} (\kappa,\varepsilon_1,\varepsilon_2
,s,t)$, {\it i.e.},
the amplitudes for given helicity $\kappa$
of $e^-$ (we neglect the electron mass and 
assume that the helicities of the electron and the positron are 
opposite) and given
polarizations of the gauge bosons, can be written in terms of
a set of $12$ linearly independent matrix elements 
\footnote{We assume CP invariance.}($6$ for each
$\kappa$), denoted by ${\cal M}_i^{\kappa}$ 
($i=1$ to $6$) with coefficient functions $F_i(s,t)$
(see, for example, Beenakker, Denner in \cite{bd}):
\begin{equation}
{\cal M} (\kappa,\varepsilon_1,\varepsilon_2,s,t) = \sum_i {\cal M}^{\kappa}_i
\left( \varepsilon_1, \varepsilon_2 , s, t \right) 
F^{\kappa}_i (s,t).
\end{equation}
The ${\cal M}^{\kappa}_i$ are the 
linearly independent Lorentz and CP-invariant objects which
can be formed from $\bar{v}$, $u$, $\varepsilon_{1,2}$ and the momenta of the 
particles. These basic matrix elements contain only kinematical 
information and the complete dependence on the polarizations. The coefficient
functions (which also have to be Lorentz and CP-invariant) contain all the
dynamics and are independent of the polarizations.

In the SM, the Born helicity amplitude for
$e^+ e^- \rightarrow W^+ W^-$ is;
\begin{equation}
{\cal M}^{SM}_{Born} = \frac{e^2}{2 s^2_W} \frac{1}{t} 
{\cal M}_1^{\kappa} \delta_{\kappa -} + e^2 \left[ \frac{1}{s} -
\frac{c_W}{s_W} g^{\kappa}_{eeZ} \frac{1}{s -M_Z^2} \right] 2 
\left( {\cal M}_3^{\kappa} - {\cal M}_2^{\kappa} \right),
\end{equation}
where the first term is from $t$-channel $\nu$ exchange 
($\delta_{\kappa -} = 1$ for $\kappa = -1/2$ and $0$ for
$\kappa = +1/2$) and the second
term is from $s$-channel $Z$ and $\gamma$ exchange. 
The electron-$Z$ coupling is
\begin{equation}
g^{\kappa}_{eeZ} = \frac{s_W}{c_W} - \delta_{\kappa -} \frac{1}{2 s_W c_W}.
\end{equation}
The ${\cal M}
_i^{\kappa}$'s are
given by
\begin{eqnarray}
{\cal M}_1^{\kappa} & = & \bar{v} (p_+) \not \! \varepsilon_1^{\ast}
\left( \not \! k_1 - \not p_+ \right) \not \varepsilon_2^{\ast}
\omega_{\kappa} u (p_-) \nonumber \\
{\cal M}_2^{\kappa} & = &  \bar{v} (p_+) \frac{\not \! k_1 - \not k_2}
{2} \left( \varepsilon_1^{\ast} \cdot \varepsilon_2^{\ast}
\right) \omega_{\kappa} u (p_-)\nonumber \\
{\cal M}_3^{\kappa} & = & \bar{v} (p_+) \left[ \not \! \varepsilon_1^{\ast}
\left( \varepsilon_2^{\ast} \cdot k_1 \right) 
- \not \! \varepsilon_2^{\ast} \left( \varepsilon_1^{\ast} \cdot k_2 \right) 
\right] \omega_{\kappa} u (p_-),
\end{eqnarray}
where $\omega _{\kappa}$ is the helicity projection operator.
We rewrite the gravity matrix element in Eqn.(\ref{mgravity1}) in terms of
the ${\cal M}_i$. After some manipulations, we get:
\begin{equation}
{\cal M}_{gravity} = \frac{4}{M_S^4} \left[ s {\cal M}_1^{\kappa}
- s (1 -\beta \cos \theta ) \left( {\cal M}_3^{\kappa} - 
{\cal M}_2^{\kappa} \right) \right],
\label{mgravity2}
\end{equation}
where $\beta = \sqrt{1 - 4 \; M_V^2 /s}$ is the velocity of the
$V$ bosons in the cm frame $(V=W$, $Z$, $\gamma$) and
$\theta$ is the angle
between $e^+$ and $V$ in the cm frame,
{\it i.e.,} the effect of the graviton exchange is to modify the coefficient
functions $F_{1,2,3}$ and no new basic matrix element is generated.

The differential cross section for unpolarized 
electrons and positrons and $W$'s is given by 
\begin{equation}
\left( \frac{d \sigma}{d \cos \theta} \right) = \frac{\beta}{128 \pi s}
\sum_{\kappa, \varepsilon_+, \varepsilon_-} \mid {\cal M}(\kappa,
\varepsilon_+, \varepsilon_-, s,t) \mid ^2 
\end{equation}
%,
with ${\cal M}(\kappa,
\varepsilon_+, \varepsilon_-, s,t) = {\cal M}^{SM}_{Born}(\kappa,
\varepsilon_+, \varepsilon_-, s,t) + {\cal M}_{gravity}(\kappa,
\varepsilon_+, \varepsilon_-, s,t)$.
To evaluate the cross sections, it 
is convenient to use the following
expressions for ${\cal M}_i$ for 
each $\kappa$, $\varepsilon_{+,-}$ in terms of $s$, $\theta$
and $M_W$ where $\theta$ is the angle 
between $e^+$ and $W^+$ in the cm frame
(these are from the appendix of \cite{bd}):
\begin{eqnarray}
{\cal M}^{\kappa}_1 (\pm,\mp) & = & 2 E^2 \sin \theta ( \cos \theta \mp
2 \kappa), 
\label{basicm1}
\end{eqnarray}
\begin{eqnarray}
{\cal M}^{\kappa}_1 (\pm,\pm) & = & 2 E^2 \sin \theta ( \cos \theta - \beta)
,\nonumber \\
{\cal M}^{\kappa}_2 (\pm,\pm) & = & 2 \beta E^2 \sin \theta,
\label{basicm2}
\end{eqnarray}
\begin{eqnarray}
{\cal M}^{\kappa}_1 (\pm,0) & = {\cal M}^{\kappa}_1 (0,\mp) 
& = \frac{\sqrt{2} E}{M_W} E^2 ( \cos \theta \mp 2 \kappa)
\left[ 2 \beta - 2 \cos \theta \mp 2 \kappa (1-\beta ^2) \right], \nonumber \\
{\cal M}^{\kappa}_3 (\pm,0) & = {\cal M}^{\kappa}_3 (0,\mp)
& =  \frac{\sqrt{2} E}{M_W} 2 \beta E^2 ( \cos \theta \mp 2 \kappa),
\label{basicm3}
\end{eqnarray}
\begin{eqnarray}
{\cal M}^{\kappa}_1 (0,0) & = & \frac{E^2}{M_W^2} 2 E^2 \sin \theta
[3 \beta - \beta^3 -2 \cos \theta],\nonumber \\
{\cal M}^{\kappa}_2 (0,0) & = & \frac{E^2}{M_W^2} 2 \beta E^2 \sin \theta
(1 + \beta ^2),\nonumber \\
{\cal M}^{\kappa}_3 (0,0) & = & \frac{E^2}{M_W^2} 8 \beta E^2 \sin \theta, 
\label{basicm4}
\end{eqnarray}
where $\pm$ denote the transverse polarizations and $0$ denotes longitudinal
polarization, $E$ is the beam energy and the matrix elements ${\cal M}
^{\kappa}_{i=
2,3}$ vanish for the combinations of the polarizations not given above
(for example, ${\cal M}_2^{\kappa}(+,-) =0$).

Radiative corrections due to virtual
$\gamma$, $Z$, $W$ exchange to the SM and graviton exchange amplitude
and hence the SM cross section are about  
$3 \% \sim O(\alpha/(4 \pi))$ (since the
corrections to the cross section are from interference with the Born
amplitude). These corrections should change
the effect of graviton exchange 
on the cross section (which is small
to begin with) by about
the same percent
since the graviton exchange effect is from interference
with the SM amplitude. \footnote{The corrections due to emission of real
soft/collinear
photons will mostly cancel in 
the ratio of the shift of
the cross section due to graviton
exchange to the SM cross section, {\it i.e.,} in the 
the percent deviation due to graviton exchange.} 
We neglect this effect. The error on the theoretical 
prediction in the SM, including radiative corrections, is
about $6$ {\em fb} for the $WW$ case which is much smaller than the
experimental (statistical) 
error at LEP2 and comparable to, but still smaller than
the statistical error
at the LC and hence we neglect this theoretical error as well. 

A similar analysis can be done for $e^+ e^- \rightarrow ZZ$. In terms of
${\cal M}^{\kappa}_i$, the SM
Born amplitude from $t$ and $u$ channel $e$ exchange is:
\begin{equation}
{\cal M}_{Born}^{SM} = \left[ \frac{e}{s_W c_W}
\left( s^2_W - \frac{1}{2} \delta_{\kappa -} \right) \right] ^2
\left[ \left( \frac{1}{t} + \frac{1}{u} \right)
{\cal M}_1^{\kappa} + 2
\frac{1}{u} \left( {\cal M}_2^{\kappa} - {\cal M}_3^{\kappa} \right) \right].
\label{zzm}
\end{equation}
The graviton exchange matrix element is given by Eqn.(\ref{mgravity2})
with \\
$\beta
= \sqrt{1 -  4 \; M_Z^2 /s}$.
The expressions for the basic matrix elements are the same as
Eqns.(\ref{basicm1}--\ref{basicm4}) with $M_W$ replaced by $M_Z$.

\section{Limits on $M_S$}
\begin{figure}
\centerline{\epsfxsize=1\textwidth \epsfbox{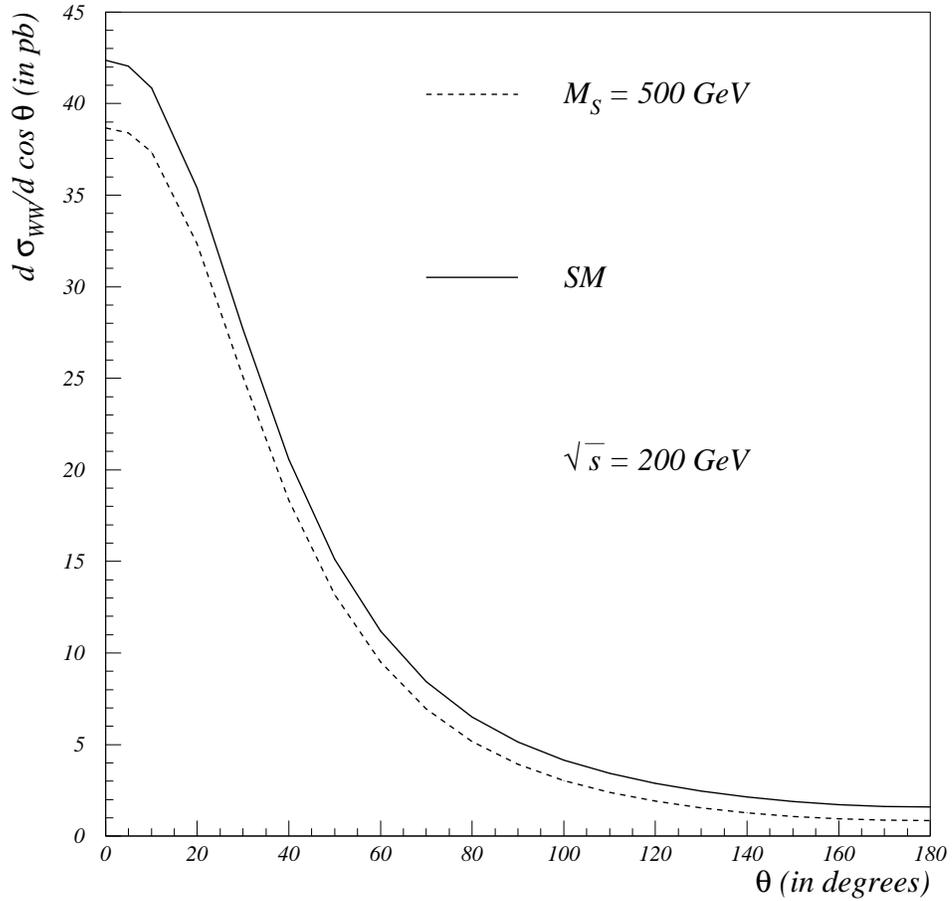}}
\caption{The differential cross section for $e^+ e^- \rightarrow
W^+W^-$ at $\protect\sqrt{s} = 200$ GeV as a function of $\theta$, the angle
between $e^+$ and $W^+$, in the SM and with the ``string scale''
$=500$ GeV.}
\protect\label{figww}
\end{figure}

In Fig.\ref{figww}, we show the differential cross section for the SM case
and with $M_S = 500$ GeV. 
The SM cross section peaks
in the forward direction due to the $t$-channel
$\nu$ exchange diagram.
We see that the effect of the graviton exchange
is to reduce the cross section for all $\theta$, if $\lambda > 0$
in Eqn.(\ref{mgravity1}) (as assumed in Fig.\ref{figww})
and to increase it for all $\theta$ if $\lambda < 0$. 
So, the deviation of the total cross
section from the SM prediction will 
give a better Confidence Level (CL) search reach for  
the graviton exchange amplitude and hence will give a better
limit on $M_S$.
The combined (all 4 detectors) LEP2 measurement at 
$\sqrt{s} = 183 - 189$ GeV is $\sigma_{WW} \approx
16$ pb $\pm 0.4$ pb \cite{lep2ww}. We get a $2 \; 
\sigma$ limit of $M_S \approx 665$ GeV assuming that the 
central value of the measurement agrees
with the SM prediction (we will make this assumption for all cases
to follow also).
For a future
integrated luminosity of $2.5$ fb$^{-1}$ combined over 4 detectors
(combining $\sqrt{s} = 162$ -- $200$ GeV) and assuming that the
error on the cross section is only statistical and the
efficiency for detecting $WW$ is $100 \%$ (it is likely
to be about $80 \%$), the future
measurement will be $\approx 16$ pb $\pm 0.08$ pb. This gives
a $2 \sigma$ limit of $M_S \approx 
1040$ GeV. \footnote{This is obtained by roughly scaling
the present $2 \sigma$ limit by a factor of (present error
on $\sigma$)$/$(future error on $\sigma$)
$^{1/4}$ since (for the same $\sqrt{s}$) the deviation in cross section
due to the graviton exchange, being mainly
due to the interference between the SM and the graviton 
exchange amplitude, scales as $M_S^{-4}$.}

\begin{figure}
\centerline{\epsfxsize=1\textwidth \epsfbox{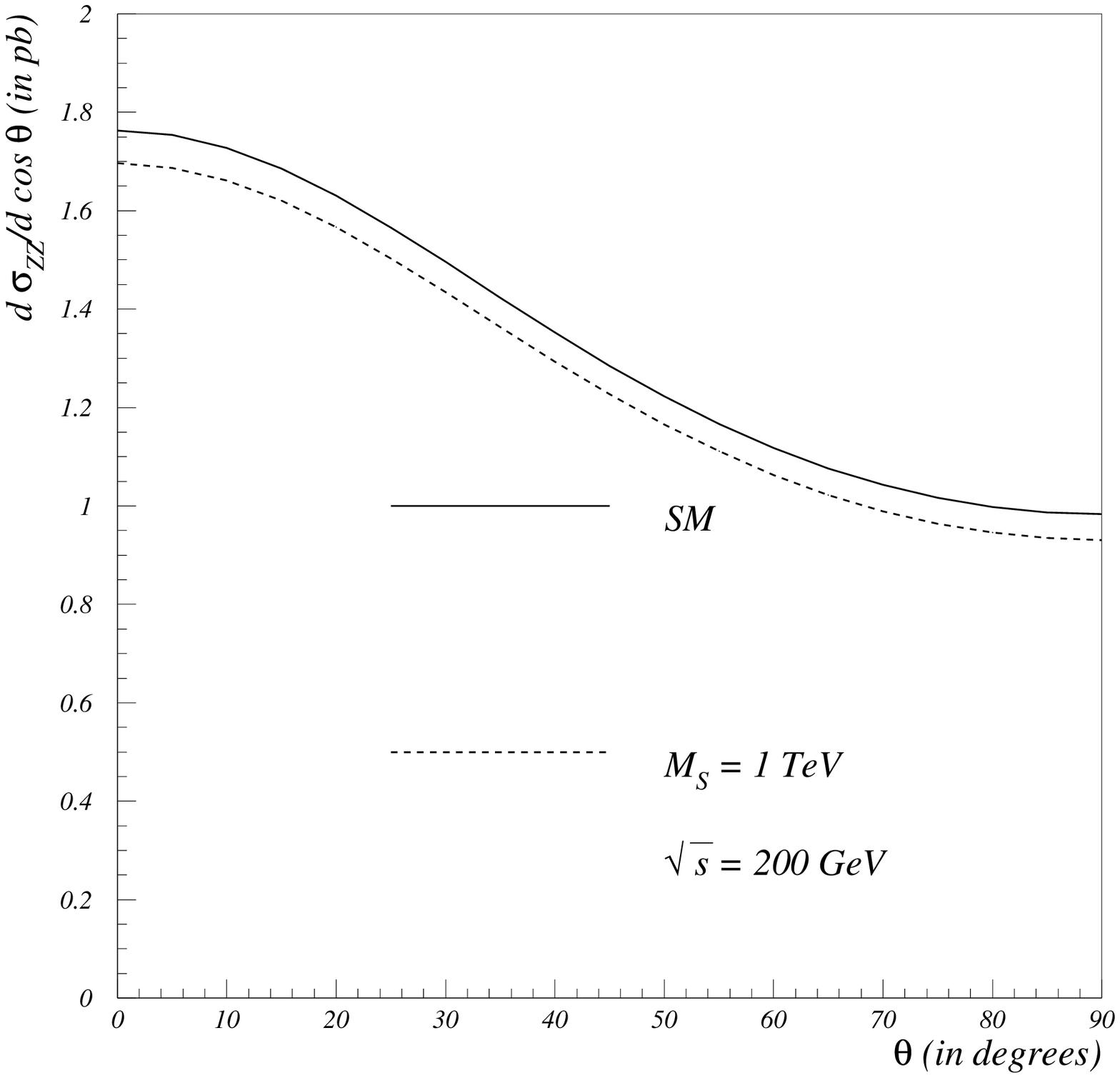}}
\caption{The differential cross section for $e^+ e^- \rightarrow
ZZ$ at $\protect\sqrt{s} = 200$ GeV as a function of $\theta$, the angle
between $e^+$ and $Z$, in the SM and with the ``string scale''
$=1$ TeV.}
\protect\label{figzz}
\end{figure}

The differential cross sections for
$e^+ e^- \rightarrow ZZ$ are plotted in Fig.\ref{figzz}.
In this case, we integrate the differential cross section
over $0 < \theta <\pi/2$ since we have identical final state particles.
The preliminary results form
the data at $\sqrt{s} = 189$ GeV
are $\sigma_{ZZ} \approx 0.6$ pb $\pm \; 0.06$
pb (we took the ALEPH error from \cite{alephzz} and divided it
by 2 for the 4 detectors assuming it is dominated by statistics).
This gives a $2 \sigma$ limit of $M_S \approx 670$ GeV.
A future measurement of $\sigma_{ZZ} \approx 0.8$ pb $\pm \; 0.02$ pb
is possible with $2$ fb$^{-1}$ summed over the 4 detectors at
$\sqrt{s} = 189$ -- $200$ GeV, assuming that the error is statistical
and the efficiency for detecting $ZZ$ is $\approx 100 \%$ (it
likely to be about $50 \%$).
This will give a $2 \sigma$ limit of $M_S \approx 980$ GeV.

At a LC with $\sqrt{s} = 500$ GeV with an integrated luminosity of
$75$ (fb)$^{-1}$, the measurements will be (with the same assumptions)
$\sigma_{WW} \approx 6.6$ pb $\pm \; 9$ {\it fb}, $\sigma_{ZZ} \approx
0.38$ pb $\pm \; 2.3$ fb. 
We get a $2 \sigma$ limits of $M_S \approx 2.8$ TeV (from $WW$) and
$3$ TeV (from $ZZ$). \footnote{At $\sqrt{s} = 500$ GeV, imposing a cut
$\cos \theta \leq 0$ 
({\it i.e.} using the effect on the ``backward'' cross section)
for WW and $\cos \theta \leq 0.9$ for $ZZ$ increases the limit
on $M_S$ by about $250$ GeV.} 

At a LC, it will be
possible to use right-handed $e^-$ beam so that the dominant
SM amplitude for
$e^+ e^- 
\rightarrow W^+ W^-$ 
due to the $t$-channel $\nu$ exchange can be suppressed.
For $M_S \sim 1$ TeV, the gravity matrix element is larger (smaller)
than
the SM matrix element for $e^- _R$ ($e^- _L$)
so that
the deviation from the SM cross section is (mainly)
due to the (square of) the
gravity matrix element for $e^- _R$ 
whereas for $e^- _L$ the effect is 
due to the 
interference between the SM and gravity amplitudes. Thus,
(assuming that the experimental error is statistical), 
for $M_S \sim 1$ TeV, the statistical significance of the deviation
in the cross section is larger for $e^- _R$ than for $e^- _L$ (or 
unpolarized $e^-$). However, for these values of $M_S$, the effect 
on the cross section with unpolarized $e^-$ is already many
standard deviations (it is an $8 \sigma$ effect even with
$1$ (fb)$^{-1}$) and so it suffices to use unpolarized $e^-$ beam.
To obtain $2 \sigma$ limits on $M_S$, we have to consider scales
$M_S \stackrel{>}{\sim} 2$ TeV for which
the graviton exchange amplitude is smaller than the SM amplitude
for {\em both} $e^- _R$ and $e^- _L$ so that the
deviation in the cross section is due 
to interference between the graviton exchange and SM amplitudes for
both beam helicities.
Therefore,  
the $2 \sigma$
limit on $M_S$
obtained by using {\em un}polarized $e^-$
is comparable to the $2 \sigma$ 
limits obtained by using polarized beams. \footnote{
For $e^- _R$ beam, the SM differential cross section peaks
in the central region ($\theta = \pi /2$) and has {\em no}
forward-backward asymmetry. The interference between the SM and graviton
exchange amplitudes (for $e^- _R$)
is constructive (destructive) in the forward (backward)
region (for $\lambda \; > \; 0$) resulting
in a forward-backward asymmetry. Thus, to get a $2 \sigma$
limit on $M_S$ using $e^- _R$
it is better to use the effect on the
forward cross section (or the forward-backward
asymmetry) rather than the total cross section.}

\begin{figure}
\centerline{\epsfxsize=1\textwidth \epsfbox{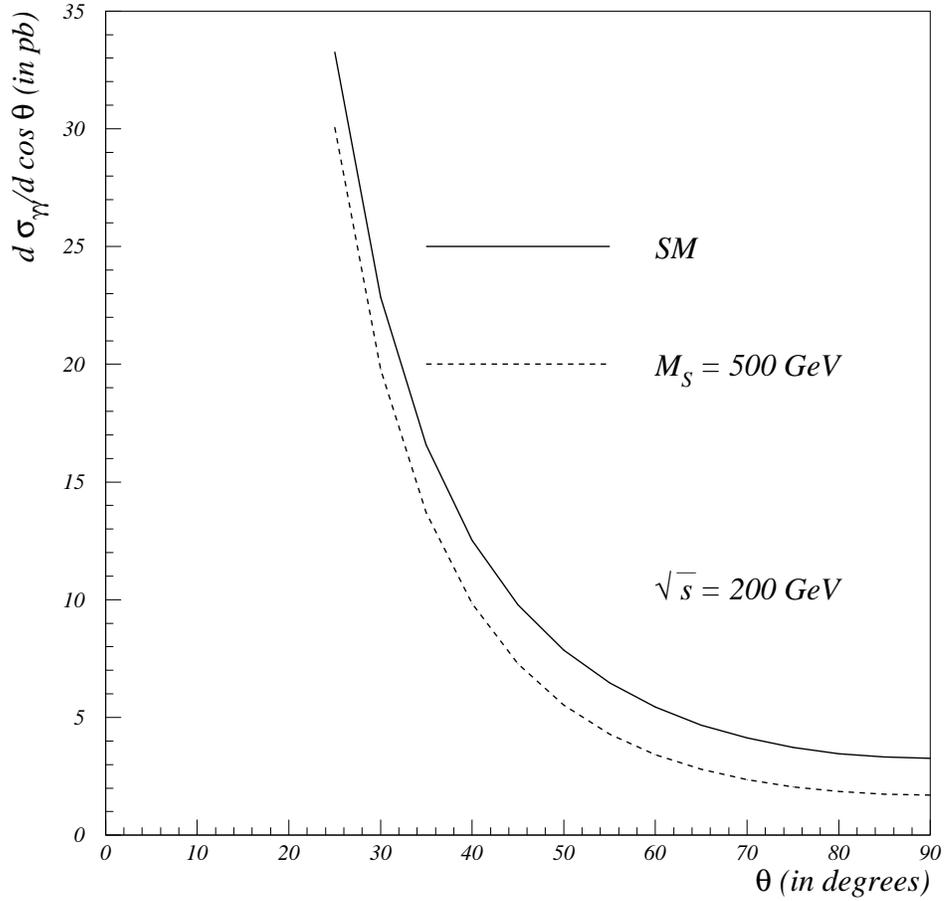}}
\caption{The differential cross section for $e^+ e^- \rightarrow
\gamma \gamma$ 
at $\protect\sqrt{s} = 200$ GeV as a function of $\theta$, the angle
between $e^+$ and $\gamma$, in the SM and with the ``string scale''
$=500$ GeV, for $\lambda = +1$ in Eqn.(\ref{mgravity1}). 
} 
\protect\label{figgg}
\end{figure}

For completeness, we show the limits
obtained by considering $e^+ e^- \rightarrow
\gamma \gamma$. 
This case for a LC with $\sqrt{s} = 1$ TeV was discussed 
by Giudice, Rattazzi, Wells in \cite{grw}.
Again, the graviton exchange matrix element
is given by Eqn.(\ref{mgravity2}) with $\beta =1$ and the
SM Born amplitude is given by Eqn.(\ref{zzm}) with the electron coupling
to $Z$ replaced by the coupling to the photon. With only
transverse polarizations for the photon, using
Eqns.(\ref{basicm1}) and (\ref{basicm2}) with
$\beta =1$, the differential cross
section simplifies to \cite{grw}:
\begin{equation}
\left( \frac{d \sigma}{d \cos \theta} \right)
= \frac{\pi}{s} \left[ \alpha G_1 
\left (\frac{t}{s} \right)  \mp 
\frac{4 s^2}{\pi M_S^4} G_2 \left (\frac{t}{s} \right)
\right] ^2
\label{sigmadiffgg}
\end{equation}
with
\begin{eqnarray}
\frac{t}{s} & = & - \sin ^2 \frac{\theta}{2}, \nonumber \\
G_1 (x) & = & \sqrt{\frac{1 + 2 x + 2 x^2}{-x (1 +x)}}, \nonumber \\
G_2 (x) & = & \sqrt{\frac{-x (1 +x) ( 1 + 2 x + 2 x^2)}{16}},
\end{eqnarray}
where the $\mp$ sign in Eqn.(\ref{sigmadiffgg})
corresponds to $\lambda = \pm 1$ in 
Eqn.(\ref{mgravity1}). 
The differential cross sections for $e^+ e^- \rightarrow \gamma \gamma$ are 
shown in Fig.(\ref{figgg}). As expected the differential cross section 
has a strong forward peak
(it diverges at $\theta = 0$ in the limit
of $m_e =0$).
The measured cross section at $\sqrt{s} = 183$ GeV is \cite{opal}
(again, we have reduced the error by a factor of 2 for the 4 detectors
assuming that the error is statistical) 
$\sigma_{\gamma \gamma} ( \cos \theta \leq 0.9 ) \approx 
8$ pb $\pm 0.2$ pb consistent
with the SM prediction giving a $2 \sigma$ limit of $M_S = 720$ GeV.
With $2.5$ (fb)$^{-1}$ of data at $\sqrt{s} = 162$--$200$ GeV, the
error will reduce to $\approx 0.06$ pb (assuming it is statistical)
giving a $2 \sigma$ limit
of $M_S = 1060$ GeV.
At a LC with $\sqrt{s} = 500$ GeV the measurement will be $
1.07$ pb $\pm 3.8$ fb giving a $2 \sigma$ limit
of $M_S = 3.2$ TeV.

\begin{figure}
\centerline{\epsfxsize=1\textwidth \epsfbox{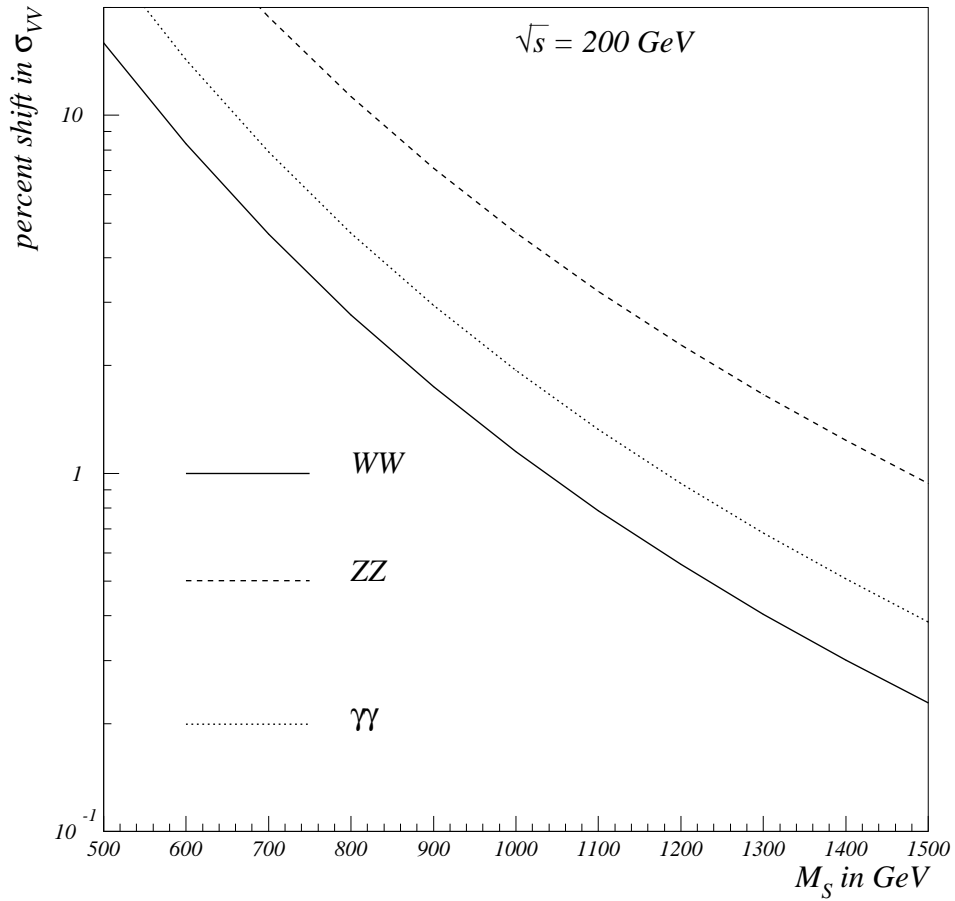}}
\caption{
The deviations in
percent of the cross section for $e^+ e^- \rightarrow W^+ W^-$,
$ZZ$, $\gamma \gamma$
from the SM prediction as a function of the ``string scale'', $M_S$
at
$\protect\sqrt{s}=200$ GeV. For the $\gamma \gamma$ case , a cut
$\cos \theta \leq 0.9$ is imposed.
}
\protect\label{figlep}
\end{figure}

\begin{figure}
\centerline{\epsfxsize=1\textwidth \epsfbox{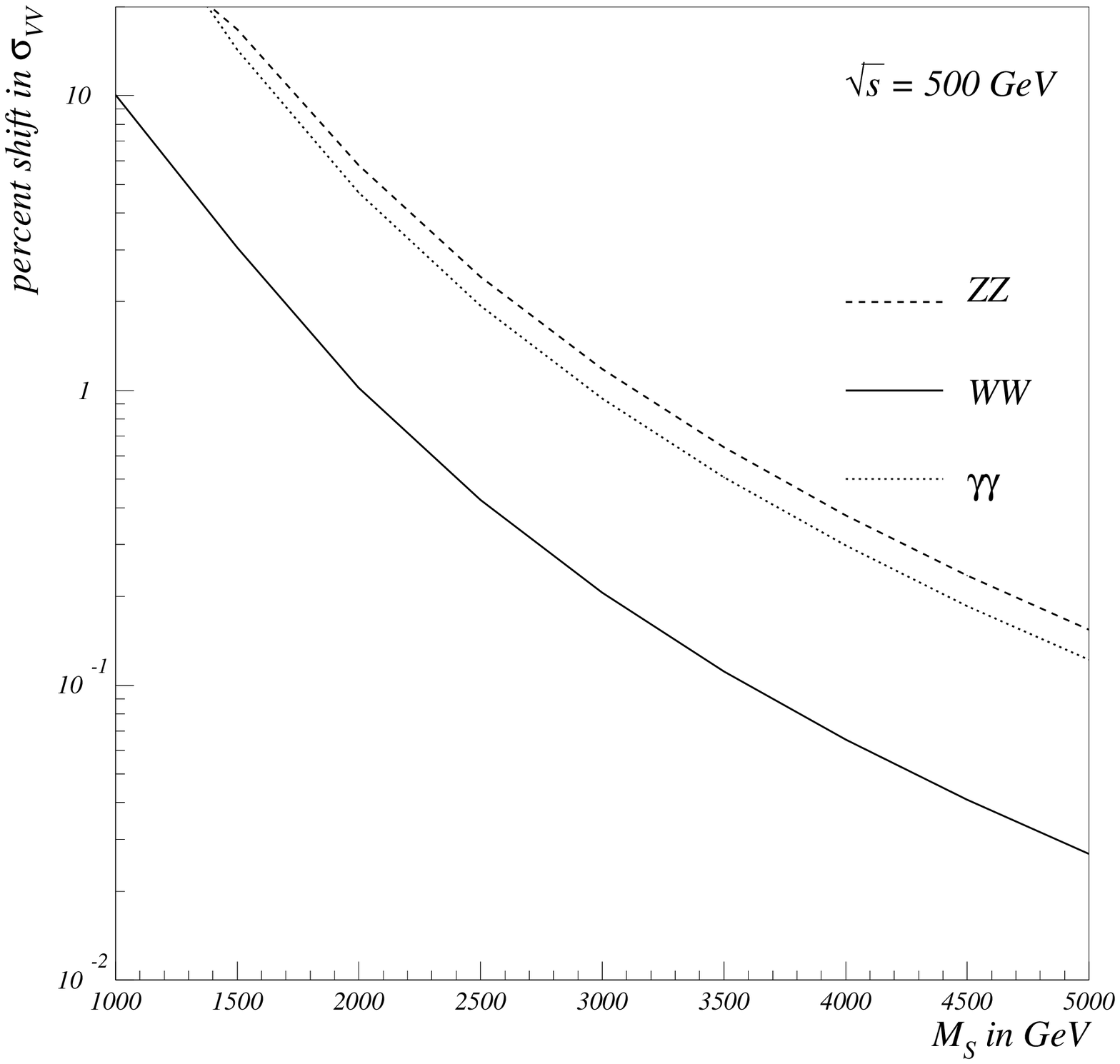}}
\caption{
The deviations in percent
of the cross section for $e^+ e^- \rightarrow W^+ W^-$,
$ZZ$, $\gamma \gamma$
from the SM prediction as a function of the ``string scale'', $M_S$
at
$\protect\sqrt{s}=500$ GeV.
For the $\gamma \gamma$ case , a cut
$\cos \theta \leq 0.9$ is imposed.}
\protect\label{figlc}
\end{figure}

To summarize, 
in Figs.\ref{figlep} and \ref{figlc}, we show the deviation from the SM cross
section for
$e^+ e^- \rightarrow WW$, $ZZ$ and $\gamma \gamma$
as a function of  
$M_S$,
the quantum gravity scale (as defined in Eqn.(\ref{mgravity1}))
for
$\sqrt{s} = 200$ GeV (LEP2) and $\sqrt{s} = 500$ GeV (LC).
We find that LEP2 with $2.5$ fb$^{-1}$ of data
will be able to put a $2 \sigma$ limit 
of
$M_S\approx 1$ TeV (corresponding to experimental
errors in the cross
sections of about $0.5 \%$, $2.5 \%$ and $0.7 \%$
for $e^+ e^- \rightarrow WW$, $ZZ$, $\gamma \gamma$,
respectively).
This is comparable to the limits from $e^+ e^- \rightarrow
f \bar{f}$ at LEP2
\cite{h} or the limits from the 
production of real gravitons at LEP2 obtained by
Mirabelli, Perelstein, Peskin in \cite{peskin}. \footnote{As mentioned 
earlier, the scale $M_S$ defined by Eqn.(\ref{mgravity1}) 
agrees with the notation
used in \cite{h}. It is related to the 
$(4+n)$-dimensional Planck scale, $M_{4+n}$ (as defined in
Eqn.(17) of the second reference in
\cite{add}) by factors (of $O(1)$) which depend on the full theory. When
comparing limits on TeV scale gravity from different processes, one should
be careful about which scale, {\it i.e.,}
$M_S$ or $M_{4+n}$ is used to derive the limit.}

\section{Acknowledgements}
We thank Bhaskar Dutta for suggestions during the beginning of this work
and David Strom for discussions
about the LEP2 measurements of $e^+ e^- \rightarrow WW$, $ZZ$ and $\gamma
\gamma$.

\end{document}